\documentclass[graybox]{svmult}

\usepackage{type1cm}        % activate if the above 3 fonts are
\usepackage{makeidx}         % allows index generation
\usepackage{graphicx}        % standard LaTeX graphics tool
\usepackage{multicol}        % used for the two-column index
\usepackage[bottom]{footmisc}% places footnotes at page bottom
\usepackage{hyperref}

\usepackage{newtxtext}       % 
\usepackage{newtxmath}       % selects Times Roman as basic font

\usepackage{natbib}

\makeindex             % used for the subject index
\begin{document}

\bibliographystyle{aps-nameyear}      % American Physical Society (APS) style, author-year citations

\title*{MOND and Methodology}
% Use \titlerunning{Short Title} for an abbreviated version of
% your contribution title if the original one is too long
\author{David Merritt}
% Use \authorrunning{Short Title} for an abbreviated version of
% your contribution title if the original one is too long
%\institute{\email{david.r.merritt@gmail.com}}

%
% Use the package "url.sty" to avoid
% problems with special characters
% used in your e-mail or web address
%
\maketitle
\vskip -1. truein

\abstract{In \textit{Logik der Forschung} (\citeyear{LDF}) and later works, Karl Popper 
 proposed a set of methodological rules for scientists. Among these were  requirements
that theories evolve in the direction of increasing content, and that new 
theories should only be accepted if some of their novel predictions are experimentally confirmed.
There are currently two, viable theories of cosmology: the standard cosmological
model, and a theory due to Mordehai Milgrom called MOND.
Both theories can point to successes and failures, but only MOND has repeatedly
made novel predictions that were subsequently found to be correct.
Standard-model cosmologists, by contrast, have almost always responded to new  
observations in a post-hoc manner, adjusting or augmenting their theory as needed 
to obtain correspondence with the facts.
I argue that these methodological differences render a comparison of the two
theories in terms of their `truthlikeness' or `verisimilitude' essentially impossible
since the two groups of scientists achieve correspondence with the facts in 
often very different ways, and
I suggest that a better guide to the theories' progress toward truth 
might be the methodologies themselves.
}

%\section{ }
%\label{Section:Intro}
\vskip 0.75 truein

Karl Popper, in his  \textit{Realism and the Aim of Science} (\citeyear{RAS}, p. 234),  identified 
two main attitudes with respect to the testing of scientific theories:

\begin{quotation}
(a) The uncritical or verificationist attitude: one looks out for `verification' or `confirmation' or
`instantiation', and one finds it, as a rule. Every observed `instance' of the theory is thought
to `confirm' the theory.
\smallskip

\noindent (b) The critical attitude, or falsificationist attitude: one looks for falsification, or for counter-instances.
Only if the most conscientious search for counter-instances does not succeed may we speak of
a corroboration of the theory.
\end{quotation}

\noindent Attitude (b) is, of course, the attitude that Popper endorsed.
A \textit{critical} scientist denies that scientific theories are verifiable.
She asserts that theories are to be judged on the basis of how well they stand up to 
critical appraisal -- to sincere attempts at refutation.

In Popper's view, accommodating a theory to known experimental or observational results does not
corroborate the theory, since ``it is always possible to produce a theory to fit any given set of
explicanda'' (Popper \citeyear{CR}, pp. 241-2); or %see also LSD p. 266
as Elie Zahar (\citeyear{Zahar1973}, p. 103) expressed it, 
``theories can always be cleverly engineered to yield the known facts.''
Corroboration occurs only when the theory survives an attempted falsification: that is: when it
predicts a previously unknown fact and that prediction is subsequently confirmed.
In the words of Imre Lakatos (\citeyear{MSRP1}, p. 38),  ``the only relevant evidence is the evidence anticipated by a theory.'' 

There currently exist at least two, viable, cosmological theories: the standard, or `concordance,'
or $\Lambda$CDM, model; and an alternative theory, the foundational postulates of which
were published by Mordehai Milgrom in 1983.
The standard model assumes the correctness of Einstein's theory of gravity and motion
(or of Newton's, in the appropriate regimes) and deals with anomalies via 
postulates relating to `dark matter' and `dark energy,' among others.
Milgrom's theory includes no dark matter (or at least, does not require it).
Observations that are explained in the standard model by invoking dark matter are explained
in Milgrom's theory by postulating a modification to Newton's (or Einstein's) laws
of gravity and motion.

Both the standard model, and Milgrom's theory -- the latter is often called `MOND', 
for MOdified Newtonian Dynamics' -- can point to  successes and failures
(McGaugh \citeyear{McGaugh2015}).
But only Milgrom's theory has repeatedly made novel predictions that were
subsequently found to be correct (\citeauthor{Merritt2020}  \citeyear{Merritt2020}).
Successes of the standard model have been, almost without exception, 
successes of accommodation:
the theory has been adjusted, or augmented, or re-interpreted in order to bring its 
predictions\footnote{Throughout this chapter I use the term `prediction' in the same way that 
Popper does, to describe a statement that follows logically (deductively) from a theory (see e.g.
item 28 in Table~\ref{Table:Rules}); it comprises `retrodiction' and `explanation' 
(e.g. Popper (\citeyear{Poverty}, p. 133)).} 
 in alignment with new observational facts, most of which constituted problems for the theory
when they first came to light.
In many cases, those new facts were predicted in advance by Milgromian researchers; 
they were `unexpected' only from the standpoint of standard-model researchers.

 Dark matter has never been detected in any experiment that 
a particle physicist would consider decisive
(\citeauthor{Liu2017} \citeyear{Liu2017}; \citeauthor{Ko2018} \citeyear{Ko2018}; \citeauthor{KisslingerDas2019} \citeyear{KisslingerDas2019})
and in the absence of such a detection
 the existence of dark matter remains an unconfirmed hypothesis.
But there is a related question that is capable of being decisively answered: 
to which methodological `school' do researchers in the two camps typically belong?
The published record betrays a distinct, and profound, difference: 
Milgromian researchers have adhered closely to Popper's methodology,
standard-model cosmologists have not.
I argue that these methodological differences render pointless
any discussion of the comparative  `truthlikeness' or `verisimilitude' of the theories 
 since the two groups of scientists achieve  correspondence with data 
 in often very different ways.
 
%%%%%%%%%%%%%%%%%%%%%%%%%%%%%%%%%%%%
\vskip 0.65 truein
\centerline{\textsection\textsection}
\bigskip	
%%%%%%%%%%%%%%%%%%%%%%%%%%%%%%%%%%%%

The typical response of a scientist to a falsifying instance -- an experiment or observation that 
contradicts a theory --  is not to discard the theory.
Scientists are more likely to retain the theory and ignore the counterexample 
(\citeauthor{SSR} \citeyear{SSR}; \citeauthor{MSRP0} \citeyear{MSRP0}).
If the refutation is persistent or compelling, the scientist may decide to tack an additional hypothesis onto the theory, 
one that targets the anomaly and `explains' it.
Ptolemy's `equants' came about in this way, as did the postulates in the standard cosmological model about `dark matter' and `dark energy' (\citeauthor{Merritt2017} \citeyear{Merritt2017}).

Karl Popper's name is most often associated with his famous demarcation criterion: the idea that scientific 
hypotheses have the quality of being falsifiable, that is, vulnerable to experimental testing.
But Popper was quite aware that scientists do not always walk away from a theory just because it has failed a test.
In fact, he wrote at length, and with considerable insight, about the methodology that scientists should follow when modifying a theory in response to refutations.

Popper's methodological guidelines were intended to preserve falsifiability:
``\textit{Only with reference to the methods applied} to a theoretical system is it at all possible to decide whether we are dealing with a conventionalist or an empirical theory'' (\citeyear{LSD}, p. 82).
He understood that, logically, falsification could always be evaded by conventionalist maneuvers: 
by ad hoc changes that simply target the anomaly.
Thus Popper required that a modified theory should do more than simply explain
 the experimental results that brought down the previous theory.
 The new theory should have more content, and it should only be accepted if it passes some new tests,
 among other requirements.

Many of Popper's methodological rules appeared in 1934, in \textit{Logik der Forschung}.
Others can be found scattered through later writings, including
\textit{Conjectures and Refutations} (\citeyear{CR}), 
\textit{Realism and the Aim of Science} (\citeyear{RAS}), and 
\textit{The Open Society and its Enemies} (\citeyear{OSI}) among others.
Jarvie \citeyearpar{Jarvie2001} lists fifteen rules; Keuth \citeyearpar{Keuth2005} finds twelve,
only six of which appear in Jarvie's list; and 
Johansson \citeyearpar{Johansson1975} compiles over twenty, including rules for the 
social sciences, the latter mostly 
from \textit{The Poverty of Historicism} \citeyearpar{Poverty}.

Table~\ref{Table:Rules} presents a concatenation of these three lists.
I have omitted the rules pertaining to social science and to probability statements, 
and I include two additional rules: no. 3 (from Popper \citeyear{LSD}, p. 253) and
no. 8 (from Popper \citeyear{CR}, p. 38).

The first rule in Table~\ref{Table:Rules} was called by Popper (\citeyear{LSD}, p. 33) the 
``supreme rule,'' that is, ``a rule of a higher type. It is the rule which says that the other rules of scientific procedure must be designed in such a way that they do not protect any statement in science against falsification''. (Keuth calls this the ``meta-rule''.)

The second rule (``The game of science is, in principle, without end'') expresses
Popper's commitment to fallibilism: the acknowledgement that we can never be certain of the 
correctness of our theories, therefore we can never stop testing them.

%%%%%%%%%%%%%%%%%%%%%%%%%%%%%%%%%%%%%%%%%%%%%%
\begin{table}[!t]
\caption{Popper's Methodological Rules$^a$}
\label{Table:Rules}
\begin{tabular}{p{0.5cm}p{9.5cm}p{2.3cm}}
\hline\noalign{\smallskip}
%
%\centering
%\begin{tabular}{c l r }
%\hline\hline
\# & Rule & Source  \\ [0.5ex]
\hline
1 & the other rules of scientific procedure must be designed in such a way that & J [SR], K [MR] \\
   & they do not protect any statement in science against falsification &  \\
2 & The game of science is, in principle, without end. He who decides one day & J1, K1 \\
   & that scientific statements do not call for any further test, and that they can be & \\
   & regarded as finally verified, retires from the game. & \\
3 & it is part of our \textit{definition} of natural laws if we postulate that they are to be & M1 \\
   & invariant with respect to space and time; and also if we postulate that they  & \\
   & are to have no exceptions & \\
4 & we are not to abandon the search for universal laws and for a coherent & J3, K4\\
   & theoretical system, nor ever give up our attempts to explain causally any & \\
   & kind of event we can describe & \\   
5 & \textit{never \ldots explain physical effects, i.e. reproducible regularities, as} & J12\\
   & \textit{accumulations of accidents}\\
6 & regard natural laws as synthetic and strictly universal statements & 3i \\
7 & only such statements may be introduced in science as are inter-subjectively & 1i, K3\\
   & testable & \\
8 & \textit{criteria of refutation} have to be laid down beforehand: it must be agreed & M2 \\
    & which observable situations, if actually observed, mean that the theory is & \\
    & refuted &\\ 
9 & in the case of a threat to our system, we will not save it by any kind of &K6  \\
   & \textit{conventionalist stratagem} & \\
10& adopt a rule not to use undefined concepts as if they were implicitly defined & 2i, J4\\
11& only those [auxiliary hypotheses] are acceptable whose introduction does & 2ii, J5, K8\\
   & not diminish the degree of falsifiability or testability of the system in  & \\
   & question, but, on the contrary, increases it & \\
12& we shall forbid \textit{surreptitious} alterations of usage & 2iii, J6, K9 \\
13& Inter-subjectively testable experiments are either to be accepted, or to be & 2iv, J7-8\\
   &  rejected in the light of counter-experiments. The bare appeal to logical  & \\
   & derivations to be discovered in the future can be disregarded. &\\ 
14& auxiliary hypotheses shall be used as sparingly as possible & 2v \\
%   & \textbf{Rules Demanding a High Degree of Falsifiability} & \\
15& the number of our axioms -- of our most fundamental hypotheses & 3iii \\
   & -- should be kept down & \\  
16& The new theory should proceed from some \textit{simple, new, and powerful,} & 3v \\
    & \textit{unifying idea} about some connection or relation (such as gravitational & \\
    & attraction) between hitherto unconnected things (such as planets and & \\
    & apples) or facts (such as inertial and gravitational mass) or new `theoretical & \\
    & entities' (such as field and particles). & \\
17& any new system of hypotheses should yield, or explain, the old, corroborated, & 3iv \\
    & regularities \\
18& those theories should be given preference which can be most severely tested \ldots  & 3ii, J11\\
    & equivalent to a rule favouring theories with the highest possible empirical content & \\ [1ex]
\hline
\end{tabular}
$^a$  `2i' indicates the i'th rule from group 2 of \cite{Johansson1975} and similarly
for J (= \citeauthor{Jarvie2001} \citeyear{Jarvie2001}) and  
K (= \citeauthor{Keuth2005} \citeyear{Keuth2005}); reference to the works by Popper
in which the rules first appeared can be found by consulting those authors.
Rules no. 3 and 8, marked `M', do not appear in any of the three lists and
references are given in the text.
\end{table}

\setcounter{table}{0}
\begin{table}[ht]
\caption{Popper's Methodological Rules (continued)}
\begin{tabular}{p{0.5cm}p{9.5cm}p{2.3cm}}
\hline\noalign{\smallskip}
\# & Rule & Source  \\ [0.5ex]
\hline
19& we require that the new theory should be \textit{independently testable} & 3vi\\
20& We shall take it [the theory] as falsified only if we discover a \textit{reproducible} & 4i \\
   & \textit{effect} which refutes the theory. In other words, we only accept the falsification & \\
   & if a low-level empirical hypothesis which describes such an effect is & \\
   & proposed and corroborated\\ 
21& we should not accept \textit{stray basic statements} -- i.e. logically disconnected & 4ii, J10 \\
   & ones -- but \ldots we should accept basic statements in the course of testing & \\
   & \textit{theories}; of raising searching questions about these theories, to be answered & \\
   & by the acceptance of basic statements & \\
22& a theory is to be accorded a positive degree of corroboration if it is & 5i \\
   & compatible with the accepted basic statements and if, in addition, a non-& \\
   & empty sub-class of these basic statements is [-- -- --] accepted as the results &\\
   & of sincere attempts to refute the theory\\
23& it is not so much the number of corroborating instances which determines & 5ii \\
    & the degree of corroboration as \textit{the severity of the various tests} to which & \\
    & the hypothesis in question can be, and has been, subjected\\   
24& we shall not continue to accord a positive degree of corroboration to a & 5iii, K12 \\
    & theory which has been falsified by an inter-subjectively testable experiment & \\
25& We require that the [new] theory should pass some new, and severe, tests. & 5iv \\
26& We choose the theory which best holds its own in competition with other & K10 \\
    & theories; the one which, by natural selection, proves itself the fittest to &\\
    & survive. This will be the one which not only \textit{has hitherto stood up to the} &\\
    & \textit{severest tests}, but the one which \textit{is also testable in the most rigorous way.} & \\
27& whenever we find that a system has been rescued by a conventionalist & K7 \\
   & stratagem, we shall test it afresh, and reject it, as circumstances may require & \\
28& With the help of other statements, previously accepted, certain singular & K5 \\
    & statements -- which we may call `predictions' -- are deduced from the [new] & \\
    & theory; especially predictions that are easily testable or applicable. From & \\
    & among these statements, those are selected which are not derivable from & \\
    & the current theory, and more especially those which the current theory & \\
    & contradicts. &\\ 
29& a theory which has been well corroborated can only be superseded by one & K11\\
    & of a higher level of universality; that is by a theory which is better testable & \\
    & and which, in addition, \textit{contains} the old, well corroborated theory -- or& \\
    & at least a good approximation to it. \\
30& Once a hypothesis has been proposed and tested, and has proved its mettle, & J2, K2 \\
    & it may not be allowed to drop out without `good reason'. A `good reason' & \\
    & may be, for instance: replacement of the hypothesis by another which is  & \\
    & better testable; or the falsification of one of the consequences of the hypothesis. & \\
31& after having produced some criticism of a rival theory, we should always & J9 \\
    &  make a serious attempt to apply this or a similar criticism to our own theory \\[1ex]
\hline
\end{tabular}
\end{table}
%%%%%%%%%%%%%%%%%%%%%%%%%%%%%%%%%%%%%%%%%%%%%%

Rules no. 3-6 enjoin the scientist to search for causal and universal laws as explanations
for observed events. 
As David Miller (\citeyear{Miller1994}, pp. 26-7) has emphasized, Popper is not implying here
any ``metaphysical assumption concerning the immutability or order of nature,''
since such an assumption is amenable to testing and may be found to be false.
Rather he is proposing the methodological rule: search for spatio-temporally
invariant laws, even though the search may turn out to be unsuccessful.

Rules no. 8-13 forbid conventionalist stratagems, that is, adjustments intended
to protect a theory from falsification.
In \textit{The Logic of Scientific Discovery}, Popper highlighted four such
``immunizing'' techniques and rules nos. 10-13 target each in turn.
\cite{Jarvie2001} separates rule no. 13 into two: admonishing scientists,
when faced with a refutation,  
not to arbitrarily reject either an experimental result or the theoretical derivation that
conflicts with it.\footnote{Both Johansson (\citeyear{Johansson1975}) and 
Jarvie (\citeyear{Jarvie2001}) note that the wording of the first part of rule no.
13 is confusing. Johansson (p. 58)  suggests that Popper meant to write
``Inter-subjectively testable \textit{theories}''; Jarvie (p. 59) suggests that ``What is
plainly intended is a presumption that inter-subjectively testable experimental work
be accepted.'' I find Jarvie's suggestion to be the more convincing.}

In addition to forbidding conventionalism, Popper proposed a number of other rules that
are relevant to theory change -- that is: to a situation in which a theory is modified
in response to a refutation.
Rules nos. 14-19, 25 and 26 together imply the following: In explaining the observations
that brought down a falsified theory, a new theory should conserve the explanatory 
successes of the old theory (rule no. 17); it should do so in a way that maximizes  
falsifiability/new content/boldness (nos. 14-16, 18);
and at least some of the modified theory's novel content should be experimentally 
corroborated (nos. 25, 26).
Of course, as stated, rule no. 25 -- that ``the theory should pass some new, and severe,
tests'' -- is not quite a \textit{methodological} rule, since, as Lakatos (\citeyear{Lakatos1968}, p. 388) 
noted, ``It is up to us to devise bold theories; it is up to Nature whether to corroborate or to refute them.''
But it is reasonable to recast the rule as a methodological one, e.g., we \textit{accept} a new theory only if it passes some new, and severe, tests.

Most scientists would probably agree with Popper about the privileged status of confirmed, novel predictions. 
For instance, Gottfried Leibniz wrote that
``Those hypotheses deserve the highest praise \ldots by whose aid predictions can be made,
even about phenomena or observations which have not been tested before'' 
(\citeauthor{Leibniz1678} \citeyear{Leibniz1678}).
Similar statements can be found in writings of John Herschel, William Whewell, Henri
Poincar\'e, Charles Peirce, Norbert Campbell and others.
But Popper makes a stronger claim.
Not only is it \textit{impressive} when a theory correctly predicts a previously unknown fact.
Popper is arguing that the confirmation of a novel prediction is the \textit{only sort of evidence that counts}.

What was the basis for this claim?
The starting point is the fallacy of induction: the logical impossibility of generalizing from discrete
instances to a general rule.
Even an incorrect theory can make correct predictions, and one can always accommodate 
a finite set of data to an infinite number of theories.

It is tempting to believe that a successful prediction always lends support to a theory,
but this belief flies in the face of the `paradoxes of confirmation' (Hosiasson-Lindenbaum\citeyear{Hosiasson1940}; Hempel \citeyear{Hempel1945}).
It is easy to show that, from a purely logical standpoint,
a universal hypothesis is supported by anything that does not contradict it; the only sort
of observation that fails to support a hypothesis is one that disproves it. 
  Thus: my observation of a red fire truck outside my window confirms the standard model of cosmology precisely
 as much (or as little) as an observation of the cosmic microwave background --
 so long as that model does
 not forbid the existence of red fire trucks.
As \citeauthor{RAS} (\citeyear{RAS}, p. 235) put it: ``Thus an observed white swan will, for the verificationist, support the theory that all swans are white; and if he is consistent (like Hempel), then he will say that an observed black cormorant also supports the theory that all swans are white.''
The `paradoxes' of confirmation are a straightforward consequence of the fallacy of induction.

Correspondence of data with theory, of itself, counts for little; 
one needs to find a sharper criterion to separate the evidentially relevant wheat from the chaff.

Popper's `positive theory of corroboration' (\citeyear{LSD}, pp. 265-73; \citeyear{RAS}, pp. 230-61) 
derives from three premises.\footnote{Philosophers who reject some or all of these premises will sometimes nevertheless embrace the \textit{conclusions} that Popper derived from them.
For instance, Psillos (\citeyear{Psillos1999}),
who makes no secret of his inductivist leanings, or of his admiration for Carnap,
writes (p. 105 and 173) ``we should not accept a hypothesis merely 
on the basis that it entails the evidence, if that hypothesis is the product of an ad hoc manoeuvre
\ldots The notion of empirical success
that realists are happy with is such that it includes the generation of novel predictions
which are in principle testable.'' 
Those sentences  could just as easily have been written by Popper 
(cf. rules no. 7, 9 and 19 from Table~\ref{Table:Rules}).
Niiniluoto (\citeyear{Niiniluoto2018}, p. 117) similarly suggests as an ``acceptance rule'' for 
an inductive inference that it
``should be independently testable, i.e. it should either explain some old evidence or be successful in serious new tests \ldots the best hypothesis is one with both explanatory and predictive power.''}
From the paradox of confirmation it follows (as just discussed) that it would be a mistake
to consider an observation as supporting a hypothesis simply because it is consistent with that
hypothesis.
The second premise was Popper's belief that, while degree of corroboration ``may at first sight
look like a probability \ldots [it] exhibits properties incompatible with the rules of the probability calculus'' (\citeyear{RAS}, p. 232).
For instance, the testability of a theory, and therefore its potential for corroboration,
 increases with its informative content, and therefore with its \textit{im}probability.
And third was Popper's insistence that a goal of science must always
be toward theories with \textit{greater} informative content -- that is: toward theories of
 lower probability.
Popper (\citeyear{RAS}, p. 222) argued that an inductivist (he singled out Carnap) 
will always try to maximize the probability of a theory
given the evidence and so will always be led to theories that go as little as possible beyond the 
evidence.
Whereas scientists, he said, ``invariably prefer
a highly testable theory whose content goes far beyond all observed evidence to
an ad hoc hypothesis, designed to explain just this evidence, and little beyond it,
even though the latter must always be more probable than the former'' (Popper
\citeyear{RAS}, p. 256).

Popper (\citeyear{OK}, p. 71) noted that 
``Knowledge never begins from nothing, but always from some background  knowledge.''
He defined background knowledge, $B$, as the set of assumptions that are 
accepted (perhaps only tentatively) when a new hypothesis is tested, and argued that
``what is interesting in a new conjecture $a$ is, in the first instance, the relative content
$a, B$; that is to say, that part of the content of $a$ which goes beyond $B$''
(Popper \citeyear{OK}, p. 49).

Based on these arguments, Popper was led
to reformulate the question `Does an observation $E$ support a hypothesis $H$?' as
`Does $E$ support $H$ in the presence of background knowledge $B$?'

Simply requiring that $E$ follow from the conjunction of $H$ and $B$ is insufficient,
since this condition may be satisfied if $E$ follows from $B$ alone.
Nor is it enough to demand that $E$ does \textit{not} follow from $B$ alone.
Popper gave as an example the failure of James Challis to
discovered Neptune, even though he was the first to observe the planet near to its
predicted location:
``The presence of \textit{some} unknown star of eighth magnitude, close to the 
calculated place, was in itself quite probable on his background knowledge and therefore
did not appear significant to him'' (Popper \citeyear{RAS}, p. 237).

These considerations led Popper (\citeyear{RAS}, p. 239) to propose that evidence $E$ 
supports hypothesis $H$ given background knowledge $B$ if both:
\begin{enumerate}
\item $E$ follows from the conjunction of $H$ and $B$
\item $E$ is improbable on the background knowledge alone.
\end{enumerate}
\noindent
Popper noted that 
saying that $E$ is improbable based on $B$ alone is similar to saying that $H$ is a bold 
hypothesis -- that it makes claims that go far beyond the background knowledge; 
and therefore that it has high empirical content.
Elsewhere, Popper had defined a ``severe test'' in essentially the same way.
Thus Popper's condition for corroboration can be stated as: A theory is corroborated
when it survives a severe test: a concerted attempt at falsification.\footnote{Miller 
(\citeyear{Miller1994}, p. 106): ``Sitting around complacently with a well-meant resolve to accept any refutations that happen
to arise is a caricature of genuine falsificationism.''}  
The more novel a test -- the more unlikely the prediction in the light of existing knowledge
-- the riskier it is, and the greater the degree of corroboration if the prediction is confirmed.

%%%%%%%%%%%%%%%%%%%%%%%%%%%%%%%%%%%%
\bigskip
\centerline{\textsection\textsection}
\bigskip	
%%%%%%%%%%%%%%%%%%%%%%%%%%%%%%%%%%%%

Subsequent authors -- while not objecting to Popper's basic reasoning -- have 
argued for different, or broader, definitions of what constitutes a `novel prediction'
or evidential support.
Elie Zahar \citeyearpar{Zahar1973} noted that Popper's criterion, which recognizes only observations made after a theory was
formulated, excludes some well-known examples from history.
For instance, the Balmer series of hydrogen was known before Bohr published his postulates
in 1913;
Kepler's laws were known to Newton; Einstein knew of the anomalous precession of Mercury's orbit. 
What matters, Zahar argued, is not the chronology so much as whether a fact ``belong[s] to the problem-situation which governed the construction of the hypothesis'' -- i.e., whether the theory was designed to explain the fact. 
On this view, a `novel fact' is one that a theory was not specifically designed to explain.

Of course, one does not always know what background knowledge was in the mind of the theorist
who designed the theory.
But there is one -- rather common -- circumstance in which it is obvious that background
knowledge is being used in this way.
That is when the theory contains unspecified parameters, and the parameters are determined from
experimental or observational data. 
In Zahar's (\citeyear{Zahar1973}, pp. 102-3) words:
\begin{quotation}\noindent
Consider the following situation. We are given a set of facts and a theory T $[\lambda_1, \dots, \lambda_m]$ which contains an appropriate number of parameters. Very often the parameters can be adjusted so as to yield a theory T* which `explains' the given facts \ldots
In such a case we should certainly say that the facts provide little or no evidential support for the theory, since \textit{the theory was specifically designed to deal with the facts}.
\end{quotation}
Zahar is not claiming here that there is anything illegitimate about determining a theory's parameters from data.
Rather, he is arguing that data that are used to set the parameters of a theory do not \textit{corroborate} the theory; they only \textit{complete} the theory; and in so doing they have lost their evidential value.
John Worrall summarized this condition more succinctly: 
 ``one can't use the same fact twice: once in the construction of a theory and then again in its support''
 (Worrall \citeyear{Worrall1978a}, p. 48).\footnote{Worrall (\citeyear{Worrall1985}, p. 313) argues further
 ``that when one theory has accounted for a set of facts by parameter-adjustment, while a rival accounts for the same facts directly and without contrivance, then the rival does, but the first does not, derive support from those facts.'' Interpreted broadly, Worrall's argument would imply that no standard-model
 explanation of any fact correctly predicted by Milgrom's theory counts in favor of the standard
 model, since standard-model explanations of such facts always invoke a multitude of adjustable parameters or auxiliary hypotheses not required by Milgrom's theory; some examples
 are discussed below.}

In practice, the situation may not be quite as clear-cut as Zahar's argument suggests.
There may be different data sets, or combinations of data sets, that a theorist can
use when determining a theory's parameters and it may not be obvious
which data should be assigned to the `background knowledge' and which data can be
considered evidentially relevant.
It is also possible that the data are not reproducible for \textit{any} choice of a theory's parameters, 
and if so they would expose the theory to potential falsification.\footnote{An example
occurred in studies of the cosmic microwave background (CMB), but the response
of standard-model cosmologists was simply to add more parameters. Early studies of the CMB
 (e.g. Jaffe et al. \citeyear{Jaffe2001}) assumed a value $n=1$ for the power-law index
 of the spectrum of initial density perturbations, but as the amount and quality of data
increased, this value $n$ began to be treated as a free parameter  (e.g. Netterfield \citeyear{Netterfield2002}) and later as a `running index' (e.g. Spergel et al. \citeyear{Spergel2007}).
In this way the model was ``immunized'' (Popper's expression) from falsification.
I am aware of only one attempt to confront the CMB data with a testable prediction;
the theory was Milgrom's and the prediction (McGaugh \citeyear{McGaughCMB1999})
was confirmed (de Bernardis et al. \citeyear{deBernardis2002}).}

But consider the following special case.
Suppose that the theory contains just one unspecified parameter and that its value is
formally over-determined by the data: that is, that there are a number of independent data
sets from which the parameter can be determined with comparable precision.
In that case, whichever data set is chosen to determine the parameter (it does not much matter which) 
becomes part of the background knowledge; the newly-determined parameter can then be inserted
into the theory and the now-completed theory can be used to make predictions which can be tested against 
the other data sets.
Those tests are novel according to Zahar's criterion, and if they are successful, the
successes constitute corroboration of the underlying theory.

%%%%%%%%%%%%%%%%%%%%%%%%%%%%%%%%%%%%%%%%%%%%%%%%%
 \begin{table}
    \begin{minipage}{338pt}
\caption[Shortened table caption for the list of tables]{Determinations of Milgrom's constant
}
\label{Table:A0Convergence}
\addtolength\tabcolsep{2pt}% to stretch columns, if required
\begin{tabular}{@{}l@{\hspace{5pt}}lll@{}}
\hline \hline
Prediction & Reference & $N_\mathrm{galaxy}$ & $a_0$ ($10^{-10}$ m s$^{-2}$) \\
\hline\\[1pt]
Baryonic Tully-Fisher Relation &\cite{Begeman1991} & $10$ & $1.21 \pm 0.24$ \\[5pt]
$(a_0 G)$		& \cite{Stark2009} & $28$  & $1.18$ \\[5pt]
		& \phantom{x} $+$ \cite{Trachternach2009} & $34$ & $1.30$ \\[5pt]
		& \cite{McGaugh2011} & $47$ & $1.24\pm 0.14$ \\[5pt]

		& \cite{LMS2016b}	& $118$ & $1.29\pm 0.06$ \\[5pt]\\[1pt]
Central Surface Density Relation & \cite{Donato2009} 	&  $\sim 10^3$      & $1.3$\\[5pt]
$(a_0/G)$	& \cite{LMSP2016} & $135$ & $1.27\pm 0.05$ \\[5pt]\\[1pt]

Radial Acceleration Relation &  \cite{WuKroupa2015}  &   $74$  &   $1.21\pm0.03$ \\[5pt]
	$(a_0/G\rightarrow a_0 G)$ &  \cite{McGaugh2016}  &   $153$    &  $1.20\pm0.02 \pm 0.24$ \\[5pt]
					&  \cite{LMSP2017}  &  &  \\[1pt]
        \\
        \hline \hline
      \end{tabular}
    \end{minipage}
\end{table}
%%%%%%%%%%%%%%%%%%%%%%%%%%%%%%%%%%%%%%%%%%%%%%%%%

This is a good description of how Milgrom's constant $a_0$
-- the only undetermined parameter in his theory -- is determined.
Table~\ref{Table:A0Convergence} (adapted from Merritt \citeyear{Merritt2020}) demonstrates that
a number of independent data sets, targeting three predictions of Milgrom's theory,
 yield comparable, and comparably accurate, estimates of $a_0$,
 approximately $1.2\times 10^{-8}$ cm s$^{-2}$.
(In fact, as discussed in the next section, one can determine $a_0$ independently using any one of hundreds of existing galaxy rotation curves, although with less precision.)
Having determined $a_0$ using any one of the data sets in Table~\ref{Table:A0Convergence},
a scientist can insert that value into Milgrom's theory and make quantitative predictions 
that are testable using any of the other data sets listed there.
As discussed in the next section, those predictions turn out to be successful; and since Zahar's criterion is satisfied for them, those successes can be said to corroborate Milgrom's theory.

The philosopher John Losee (\citeyear{LoseeProgressBook}, p. 156; 
\citeyear{LoseeScrapHeapBook}, p. 166)
uses the term `convergence' to describe cases like this: in which 
a new constant of nature is determined, consistently, from a number of different kinds of data.
Losee writes ``I know of no plausible countercase in which convergence of this kind is achieved in the case of a transition judged not to be progressive on other grounds'':
\begin{quotation}\noindent
The convergence of various determinations of the value of Avogadro's number on $6.02 \times 10^{23}$ molecules/ gram molecular weight warrants as progressive the transition from theories of the macroscopic domain to the atomic-molecular theory of its microstructure. 
And the convergence of various determinations of the value of Planck's constant on 
$6.6 \times 10^{-27}$ erg-sec warrants the transition from classical electromagnetic theory to the theory of the quantization of energy (Losee \citeyear{LoseeProgressBook}, pp. 156-7).
\end{quotation}

\noindent Losee (who was unaware, apparently, of Milgrom's theory)  adds 
``Unfortunately,  opportunities to apply the convergence condition are rare within the history of science.''

Does the standard cosmological model provide any opportunities for testing convergence?
Indeed it does: the mean baryon\footnote{Standard-model cosmologists often use  
`baryonic matter' to mean `normal [i.e. non-dark] matter'.
Milgromian researchers sometimes follow suit, even though, from their perspective,
there is no need to distinguish between two sorts of matter.} density, $\rho_b$, is a parameter that can be measured in a number of independent
ways.
It is traditional to express this quantity in terms of the dimensionless `concordance' parameter $\Omega_b$ as
\begin{equation}
\label{Equation:DefineRhoB}
\rho_{\mathrm{b}} = \frac{3}{8\pi G} \Omega_b H^2
\end{equation}
with $H$ the Hubble (expansion) parameter.
The concordance value of $\Omega_b h^2$ is said to be $0.022$ where $h\equiv H_0/100$ km s$^{-2}$.
Prior to observations of the CMB in the early 2000s, the value of $\rho_b$ was determined from two,
quite different sorts of data: (\textit{i}) the measured abundance of $^7\mathrm{Li}$ in the atmospheres of Population 
II stars in the halo of the Milky Way, together with the equations of big-bang nucleosynthesis; 
and (\textit{ii}) direct census of the density of matter in the local universe.
Both techniques yielded (and continue to yield) $0.011 \lesssim \Omega_bh^2\lesssim 0.016$ --consistent 
with each other,  and roughly one-half of the current concordance value.
Since about 2002, standard-model cosmologists have \textit{excluded} these two data sets when 
determining the values of the parameters that define their `concordance' model; 
the resulting discrepancies
in the value of $\rho_b$ are called by them the `lithium problem' 
(e.g. Fields \citeyear{Fields2011})
and the `missing baryons problem' (e.g. Shull et al. \citeyear{Shull2012}).
Thus, the evolution of the standard cosmological model beginning around 2002 violated rules
no. 17 and 29 in Table 1: it failed to conserve ``the old, corroborated, regularities'' of the model,
namely, the convergence of measured values of $\rho_b$.

Neither the `lithium problem' nor the `missing baryons problem' exists from the 
standpoint of a Milgromian researcher,
who is likely to prefer the value of $\rho_b$ that was established prior to 2000.

%%%%%%%%%%%%%%%%%%%%%%%%%%%%%%%%%%%%
\bigskip
\centerline{\textsection\textsection}
\bigskip	
%%%%%%%%%%%%%%%%%%%%%%%%%%%%%%%%%%%%

Milgrom's theory is a response to 
an anomaly that arose in the 1970s in observations of disk galaxies.
The speed, $V$, at which stars or gas clouds orbit at distance $R$ about the galaxy center
is predictable using Newton's laws of gravity and motion given the observed distribution of 
mass (`baryons') in the galaxy. 
The Newtonian prediction is often found to be reasonably correct near the centers of galaxies, i.e. $V_\mathrm{obs}(R) \approx V_\mathrm{Newton}(R)$.  
But at sufficiently large $R$, rotation curves become `asymptotially flat': 
the orbital speed tends to a constant value (different in different galaxies), 
$V_\mathrm{obs}(R) \rightarrow V_\infty \gg V_\mathrm{Newton}$.

In his first paper from 1983, \citeauthor{Milgrom1983a} proposed a modification to Newton's
laws of gravity and motion that targets, and explains, the asymptotic flatness of galaxy rotation curves.
Milgrom's auxiliary hypothesis was presented in the form of three postulates, which were re-stated,
in slightly different form, in two subsequent papers from the same year.
 For the sake of definiteness I will take the liberty of (re-)stating Milgrom's postulates as follows:
 
 %%%%%%%%%%%
 \begin{enumerate}
\item Newton's second law relating acceleration to gravitational force is asymptotically correct when applied to motion for which the gravitational acceleration is sufficiently large, but breaks down when the  acceleration is sufficiently small.
 
\medskip
\item In the limit of small gravitational accelerations, the acceleration of a test particle, in a symmetric and stationary gravitating system, is given by $\left(a/a_0\right)\boldsymbol{a} \approx \boldsymbol{g}_\mathrm{N}$, where $\boldsymbol{g}_\mathrm{N}$ is the conventional gravitational acceleration and $a_0$ (`Milgrom's constant') is a constant with the dimensions of acceleration. 

\medskip
\item The transition from the Newtonian regime to the low acceleration regime is determined by Milgrom's constant.
The transition occurs within a range of accelerations of order $a_0$ around $a_0$.

\end{enumerate}

\noindent
Sufficiently far from the center of a galaxy, the Newtonian gravitational acceleration has magnitude
$\left|\boldsymbol{g}_\mathrm{N}\right| \approx GM_\mathrm{gal}/r^2$, 
with $M_\mathrm{gal}$ the total mass of the galaxy and $r$ the distance measured from the galaxy center. 
Milgrom's second postulate, $(a/a_0)\boldsymbol{a} \approx \boldsymbol{g}_\mathrm{N}$, 
then implies

\begin{equation}
\label{Equation:aofM}
a \approx \left(a_0 GM_\mathrm{gal}\right)^{1/2}r^{-1}   .							
\end{equation}

\noindent Equating this expression with the centripetal acceleration of a test mass moving in a circular orbit of
radius $r=R$, or $V^2/R$, yields

\begin{equation}
\label{Equation:BTFR}
\frac{V^2}{R} = \frac{\left(a_0 GM_\mathrm{gal}\right)^{1/2}}{R} \ \ \ \  \mathrm{i. e.} \ \ \ \  
V = \left(a_0 GM_\mathrm{gal}\right)^{1/4}  ,
\end{equation}

\noindent so that $V$ is independent of $R$. 

As Milgrom pointed out in the same three papers from 1983, 
\nocite{Milgrom1983b,Milgrom1983c} his postulates
can be used to generate additional, testable predictions. 
As we will see, most or all of these predictions have been observationally confirmed.
But before discussing those results we can pause and take stock of how well Milgrom's
proposed changes to Newton's laws accord with Popper's methodological rules:

Rules no. 7, 11 and 19, which require that theory modifications be testable and that
they result in increased empirical content, are (as just noted) clearly satisfied.
Rule no. 8 (``\textit{criteria of refutation} have to be laid down beforehand'')
is also clearly satisfied: it is obvious from Milgrom's 1983 papers that he viewed
his predictions as having the potential to falsify the underlying theory.

Rules no. 14 and 15, which counsel parsimony when adding axioms (three in this case), 
are arguably satisfied, as is rule no. 16 (``should proceed from some simple, new,
and powerful'' idea): a proposal that Newton's laws are incorrect is nothing if not 
``new, and powerful.''
And Milgrom's auxiliary hypotheses clearly preserve ``old, corroborated, regularities''
(rule no. 17): both in the high-acceleration regime ($a\gg a_0$) since the modified
theory makes the same prediction as Newton's laws; and also in the low-acceleration
regime, since the only ``regularity'' that was known to exist in this regime ca. 1980
was the asymptotic flatness of rotation curves.
(In other words: no `Kuhn losses' here.)
 None of the other rules 1-24 (to the extent that they are applicable) is violated.

This brings us to rule no. 25, which demands that the modified theory ``should pass some new, and severe, tests'' -- in other words, that (at least some of) its novel predictions should be experimentally corroborated.
And here, Milgrom's modified dynamics has performed not just adequately, but -- by any
reasonable standard -- spectacularly.
Here is a partial list of corroborated novel predictions:

\begin{enumerate}

\item A universal relation between asymptotic speed and total mass of a disk galaxy

\item A universal relation between the acceleration $\boldsymbol{a}$ at any point in a disk galaxy and the Newtonian gravitational acceleration $\boldsymbol{g}_\mathrm{N}$ due to the galaxy's mass

\item A universal relation between the central surface densities of normal and `dark' matter in galaxies 

\item A predicted dependence of the rms, vertical velocity of stars on distance above or below the plane of the Milky Way galaxy

\item A relation between the mass of a gravitating system and the root-mean-square velocity of its components

\end{enumerate}

%%%%%%%%%%%%%%%%%%%%%%%%%%%%%%%%%%%%%%%%%%%%%%%%%
\begin{figure}
\label{Figure:BTFR}
\centering
\begin{minipage}{.5\textwidth}
  \centering
  \includegraphics[width=0.84\linewidth]{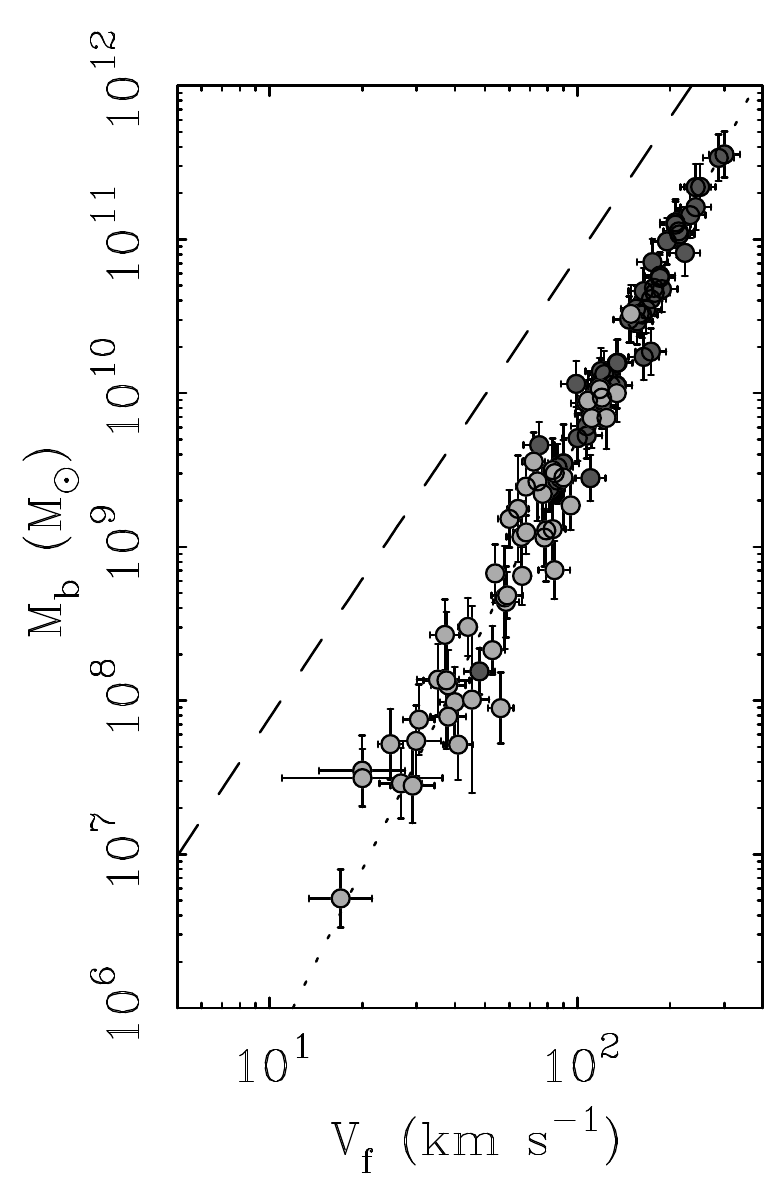}
%  \label{Figure:BTFR}
\end{minipage}%
\begin{minipage}{.5\textwidth}
  \centering
  \includegraphics[width=0.86\linewidth]{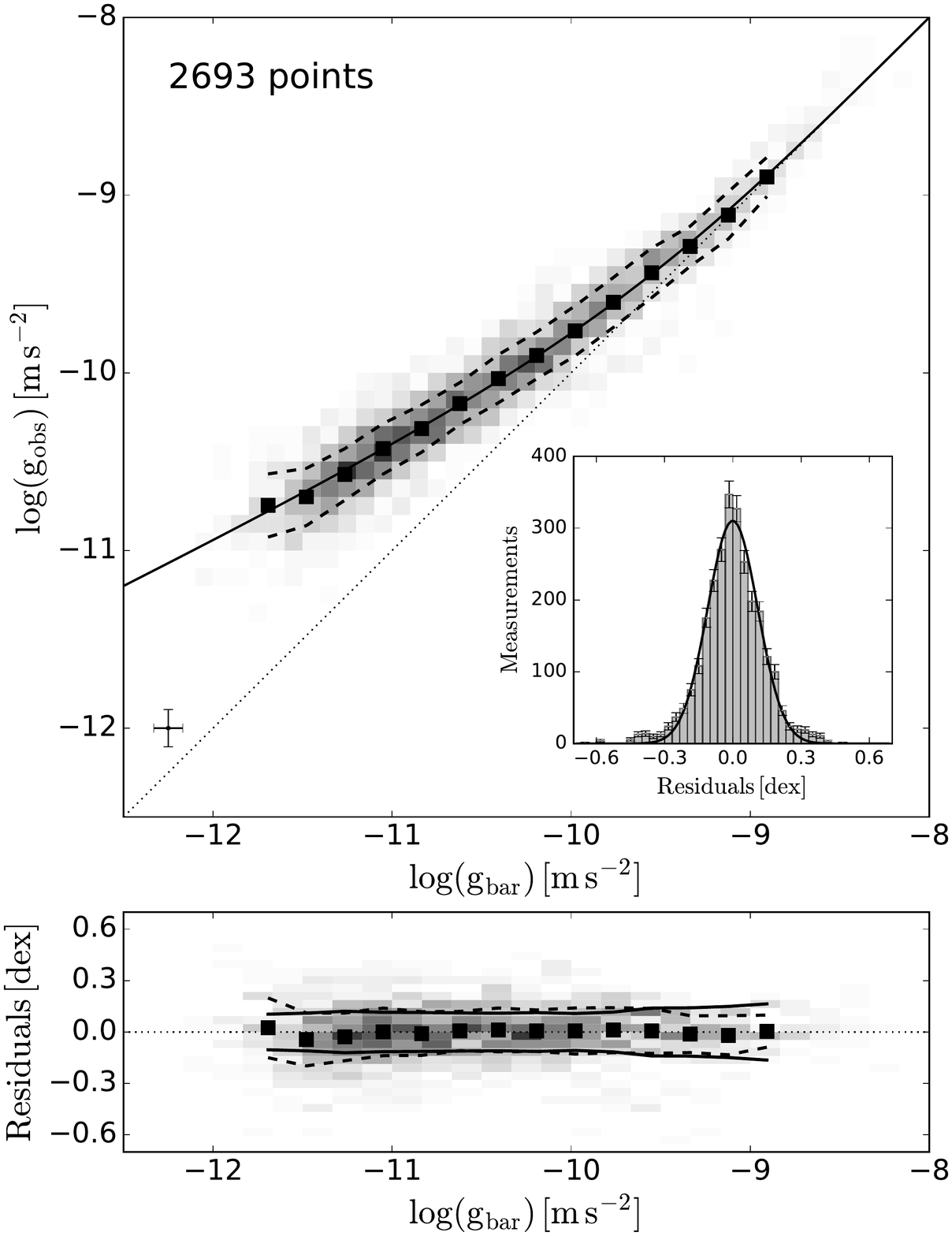}
%  \label{Figure:RAR}
\end{minipage}
\caption{Two confirmed, novel predictions of Milgrom's theory.
\textit{Left}: The baryonic Tully-Fisher relation (BTFR).
Each point corresponds to a single disk galaxy.
The vertical axis (M$_\mathrm{b}$) is the summed mass in stars and gas (the subscript `b' stands for `baryonic,' i.e.,  `non-dark').  
The horizontal axis (V$_\mathrm{f}$) is the outer, flat rotation velocity as inferred from 21 cm radio telescopic observations, what is called here `$V_\infty$'. 
Figure reprinted with permission from B. Famaey and S. S. McGaugh, ``Modified Newtonian dynamics (MOND): 
observational phenomenology and relativistic extensions,'' \textit{Living Reviews in Relativity}, 15, 2012, p. 20.
\textit{Right}:
The radial acceleration relation (RAR), derived from rotation curve data of 153 galaxies. 
The vertical axis plots the observed acceleration, $g_\mathrm{obs} = V^2/r$, or what is called $a$ in the text. 
The horizontal axis plots $g_\mathrm{bar} = \left|\partial\Phi/\partial r\right|$, or what is called $g_\mathrm{N}$ in the text.
Figure reprinted with permission from S. S. McGaugh, F. Lelli, and J. M. Schombert, 
``Radial acceleration relation in rotationally supported galaxies,''
\textit{Physical Review Letters}, 117, p. 201101, 2016.  Copyright (2016) by the American Physical Society.}
\end{figure}
%%%%%%%%%%%%%%%%%%%%%%%%%%%%%%%%%%%%%%%%%%%%%%%%%

Prediction no. 1 is just Eq.~(\ref{Equation:BTFR}). 
The \textit{observed} relation between galaxy mass and asymptotic rotation speed
(Fig.~1a) is nowadays called the
`baryonic Tully-Fisher relation' or BTFR.\footnote{That rather baroque name is due
to standard-model cosmologists; see e.g. Merritt (\citeyear{Merritt2020}, Chapter 4)
for the relevant history.
Milgrom refers to his predicted relation by the much more apt name
 `mass--asymptotic speed relation.'}
The \textit{predicted} relation contains (like essentially all predictions from Milgrom's theory)
the unspecified constant $a_0$, `Milgrom's constant.'
As noted earlier (cf. Table~\ref{Table:A0Convergence} and the accompanying discussion), 
there are many ways to determine $a_0$ from data but most
astrophysicists consider the BTFR to be the `cleanest,' that is, least subject to systematic 
errors.
The novelty of Milgrom's prediction can perhaps best be attested by the fact that
no standard-model cosmologist had proposed (or, it appears, searched for) 
any such relation prior to 1983 -- no  doubt in large part because, under the standard model, 
the asymptotic velocity is attributable almost entirely to the dark matter, not the `baryons.'

Prediction no. 2 is probably the most remarkable, at least from the standpoint of 
standard-model expectations.
\cite{Milgrom1983a} showed that the local acceleration $a$ -- accessible,
in any disk galaxy, through $V^2(R)/R$ -- must be related to the gravitational
acceleration $g_\mathrm{N}$ computed from the observed mass distribution
under Newtonian gravity via a relation $a = f(g_\mathrm{N}/a_0)g_\mathrm{N}$  
with $f(x)$ some as yet unspecified function of $x\equiv g_\mathrm{N}/a_0$. 
(Of course, in the purely Newtonian case, $f\equiv 1$.)
Regardless of the functional form of $f$, Milgrom's prediction was therefore that the acceleration
 will be a \textit{universal function of the Newtonian prediction, hence of the baryonic 
 mass distribution, with $a_0$ as the only unspecified parameter}.
 As noted by \cite{Milgrom2016c}, this prediction can be expressed in a number of essentially
 equivalent ways: as a relation between $a$ and $g_\mathrm{N}$, between $a$ and $a/g_\mathrm{N}$, 
 between $g_\mathrm{N}$ and $a/g_\mathrm{N}$ etc. 
 And in fact, observational corroboration of the prediction exists in three forms: individual rotation curves;
 the `mass discrepancy-acceleration relation' (MDAR), i. e. the relation between $a\equiv V^2/R$ and 
 $a/g_\mathrm{N}$; and the `radial-acceleration relation' (RAR), the relation between $a$ and $g_\mathrm{N}$ (shown in Fig.~1b).
 
 Given data like those in Fig.~1b, one can simply `read off'
 (modulo measurement uncertainties) the functional form of the relation between
 $a$ and $g_\mathrm{N}$.
 The latter is sometimes called the `transition function.'
 Various ad hoc, analytic forms have been proposed for the transition function; the curve 
plotted in Fig.~1b
has a form first suggested by \cite{McGaugh2008},
\begin{equation}
\label{Equation:McGaughnuofy}
f(y) = \left[1-\exp\left(-y^{1/2}\right)\right]^{-1}, \ \ \ y \equiv \frac{a}{g_\mathrm{N}} .
\end{equation}
Note the remarkable fact that Eq.~(\ref{Equation:McGaughnuofy}) (to the extent
that it is accurate) then allows one to predict the acceleration experienced by
a test body orbiting \textit{anywhere} in a galactic disk -- not just in the `asymptotic'
regimes of high or low acceleration
Thus, for instance, one can predict the rotation curve $V(R)$ for any single disk galaxy
having a well-determined mass distribution and there are dozens of published 
examples of this kind (Milgrom~\citeyear{Milgrom1988} being one of the earliest).
Perhaps the most striking of these studies (e.g. de Blok and McGaugh \citeyear{deBlokMcGaugh1997})
are based on galaxies which a standard-model
cosmologist would claim are `dark matter dominated' everywhere -- that is: for which the observed
rotation speed greatly exceeds the Newtonian prediction at all positions.
A standard-model cosmologist would predict that the rotation curve of such a galaxy
is essentially independent of the (non-dark) matter distribution, and yet one finds, in every
case, that the rotation curve is correctly predicted by the modified dynamics using only the
observed density in stars and gas.

Given the background knowledge that existed ca. 1980,
the proposal that the kinematics of any disk galaxy could be predicted, with high accuracy,
from the observed distribution of normal matter alone was amazingly bold.
There was simply no basis, under the standard model, for believing any such thing, and yet
it turned out to be correct.

The remaining predictions no. 3-5 and their observational corroboration are discussed in detail
in Merritt (\citeyear{Merritt2020}).

A number of Popper's remaining methodological rules 26-31 are expressed with
reference to the competing theory, to which I now turn.
In what follows I will restrict the discussion to predictions that follow from the
non-relativistic versions of both theories; see Merritt (\citeyear{Merritt2020}) for
a discussion of the relativistic theories.

%%%%%%%%%%%%%%%%%%%%%%%%%%%%%%%%%%%%
\bigskip
\centerline{\textsection\textsection}
\bigskip	
%%%%%%%%%%%%%%%%%%%%%%%%%%%%%%%%%%%%

Standard-model cosmologists deal with the rotation curve anomaly 
by leaving Newton's laws intact and adding an auxiliary hypothesis:\footnote{Few 
cosmology textbooks acknowledge that the existence of dark matter is a postulate.
Standard-model cosmologists take it for granted, apparently, that the existence 
of dark matter has been verified by rotation curve studies; 
e. g. Schneider (\citeyear{Schneider2015}, p. 77):
``The rotation curves of spiral galaxies are flat up to the maximum radius at which they can be measured; \textit{spiral galaxies contain dark matter}'' (italics his). 
Milgrom deserves credit for emphasizing that
the existence of `dark matter' is a postulate of the standard model
and not a confirmed fact.}
\begin{description}[SCM1]
\item[SCM1]{In any galaxy or galactic system for which the observed motions are inconsistent with the predictions of Newton, the discrepancy is due to gravitational forces from dark matter in and around the galaxy(ies).}
\end{description}
\noindent This auxiliary hypothesis has far less informative
content -- that is, it is far less \textit{testable} -- than Milgrom's.
As we have seen, Milgrom's hypothesis allows one to predict 
the rotation curve of a disk galaxy given measurements of the density of the disk.
That prediction is easily testable using the observed motions of stars or gas clouds in the disk,
and this is true for any of the (hundreds or thousands) of well-observed disk galaxies.

The standard-model hypothesis SCM1 makes no prediction about the behavior of 
\textit{observable} matter.
It only says something about the \textit{dark} matter.
An observed rotation curve can not be used to test hypothesis SCM1;
rather, that hypothesis instructs the scientist to \textit{use} the measured $V(R)$ 
(together with the observed density of stars and gas in the disk) to predict the dark matter distribution.
The rotation curve is treated as part of the background knowledge.
In Milgrom's (\citeyear{Milgrom1989a}, p. 216) words, postulate SCM1
 ``simply states that dark matter is present in whatever quantities and space distribution is needed to explain away whichever mass discrepancy arises.''
 Worrall's rule: ``One can't use the same fact twice: once in the construction of a theory and then again in its support'' tells us that the rotation curve, by virtue of having been used in
 the construction of the dark matter distribution, has lost its ability to provide
 support to the dark matter hypothesis.

This lack of testability extends to cases in which a galaxy's rotation curve is
supplemented by other kinds of kinematical data.
One much-discussed example is the so-called `Oort problem': understanding what the
observed distribution of stellar velocities \textit{perpendicular} to the Milky Way disk
(at the Solar circle) implies about the local mass density in the disk.
Milgrom's theory makes a solid prediction (this is prediction no. 4 in the list of the previous
section) and that prediction has been confirmed: that is, the vertical motions
 are observed to be consistent, under the modified dynamics, with the observed 
 (`baryonic') mass in the disk
(Nipoti et al. \citeyear{Nipoti2007b}; Bienaym\'e et al. \citeyear{BienaymeFamaey2009}).

Under the standard model, one could imagine using SCM1, together with the observed
Milky Way rotation curve, to estimate the dark matter density near the Sun; then 
test whether the vertical force generated by the total local density (`baryonic' plus dark) 
correctly predicts the observed vertical motions.
But standard-model cosmologists have never succeeded in doing this.
In the first such studies, John Bahcall (\citeyear{Bahcall1984a}; \citeyear{Bahcall1984b})
assumed a spherical dark `halo' having a mass distribution designed 
to explain the Galaxy's rotation curve.
That model made a definite prediction about the local dark matter density, as well as its
dependence on distance above or below the disk.
But Bahcall recognized the degeneracy of the dark matter models:

\begin{quotation}
Since we haven't yet observed the unseen material, we don't know how it is distributed.  Therefore we have to try different models for the unseen material to see how the results depend upon our assumptions \citep[p. 19]{Bahcall1987}.
\end{quotation}

\noindent And indeed he treated the local dark matter density and its dependence on $z$ as 
adjustable quantities.\footnote{Although Bahcall never claimed to be testing a standard-model prediction, 
he did note (Bahcall \citeyear{Bahcall1987}) that the data were explainable via Milgrom's theory.}
Subsequent studies of the Oort problem by standard-model cosmologists 
 have likewise shied away from casting their analyses as tests.
For instance, \cite{Smith2012} write (italics added):

\begin{quotation}
If we assume our background mass represents the dark halo, it corresponds to a local dark matter density of 0.57 GeV cm$^{-3}$, which is noticeably larger than the canonical value of 0.30 GeV cm$^{-3}$ typically assumed \ldots As pointed out by various authors  \ldots ,
\textit{the local dark matter density is uncertain by a factor of at least two}. Our analysis adds still more weight to the argument that the local halo density may be substantially underestimated by the canonical value of 0.30 GeV cm$^{-3}$ 
(Smith et al. \citeyear{Smith2012}, p. 11).\footnote{$0.30$ GeV cm$^{-3}  \approx 0.008 M_\odot$ pc$^{-3}$.}
\end{quotation}
\noindent
The local value
of $\rho_\mathrm{DM}$ -- the prediction that could be refuted via an analysis of the 
vertical motions -- is typically decoupled from the rotation curve constraint in these
studies by allowing the dark matter halo to be nonspherical (e.g. Garbari et al.
\citeyear{Garbari2012}; Bienaym\'e et al. \citeyear{Bienayme2014}). By treating the halo axis ratio as an extra, freely adjustable parameter, many values of the local dark matter density can be made
consistent with a given rotation curve, thus effectively nullifying any predictive power of SCM1.
(This degeneracy adds to the ``factor of at least two'' uncertainty mentioned in the quotation
from Smith et al.)
And indeed some studies are forced to assume extremely contrived shapes for the 
dark matter halo in order to get the vertical kinematics `right' (Read \citeyear{Read2014}).

One way to make a \textit{testable} prediction of the standard-model postulate
is to couple SCM1 with some other hypothesis.
In fact, standard-model cosmologists routinely assume, in addition to SCM1, that
\begin{description}[SCM2]
\item[SCM2]{The dark matter of SCM1 is composed of elementary particles.}
\end{description}
(E.g.  Funk (\citeyear{Funk2015}, p. 12264): ``Today, it is widely accepted that dark matter 
exists and that it is very likely composed of elementary particles, which are weakly interacting and massive.'')
One novel prediction immediately follows: some of the dark particles associated with the 
Milky Way must be passing at every moment through an Earth-based laboratory
and could be detected, through their interaction with normal matter.
But this prediction -- while capable of being confirmed -- is not refutable, since nothing whatsoever
is known about the properties of the putative particles (aside from the fact that no \textit{known}
particles have properties that would make them acceptable candidates).
A failure to detect the particles might simply mean that their cross-section for interaction with the
normal matter in the detectors is very small, and that is in fact one explanation that
particle physicists propose for their almost four-decade failure to detect a signal
(Bertone and Hooper \citeyear{BertoneHooper2018}).\footnote{
When Elena Aprile was asked to estimate a cost for her XENONnT dark matter experiment at the
 Italian Gran Sasso National Laboratory, the \textit{New York Times} reports that she
``was reluctant to put a price on the project. An earlier version of the experiment with 3.3 tons of xenon cost \$30 million. But that didn't include the people, she said. A big part of the cost is xenon itself, which costs around \$2 million per ton, she added. Her new detector will have 8.5 tons''
(Overbye \citeyear{Overbye2020}).
There are about a half-dozen such experiments currently underway
(as reviewed by Kisslinger and Das \citeyear{KisslingerDas2019}).
Given that the hypothesis being tested by the direct-detection experiments 
(that dark particles are passing through the laboratory) is not refutable, it is reasonable to 
ask what will have been accomplished by those experiments assuming 
the continued absence of a detection.}

In this respect, the particle dark matter hypothesis is in a state similar to that of the atomistic hypothesis at the end of the 19th century. 
Popper (\citeyear{RAS}, p. 191) noted that the hypothesis that atoms exist was, for a long time, too vague to be refuted:
\begin{quotation}
Failure to detect the corpuscles, or any evidence of them, could always be explained by pointing out that they were too small to be detected. Only with a theory that led to an estimate of the size of the molecules was this line of escape more or less blocked, so that refutation became in principle possible.
\end{quotation}
Popper's statement is perfectly applicable to the (particle) dark matter hypothesis if one replaces 
`size of the molecules' by `cross section of interaction of the dark particles with normal matter.'

We are now in a position to assess how well postulates SCM1 and SCM2 accord with
Popper's methodological rules.
Rules no. 7, 11 and 19 are violated: as we have seen, the postulates have little if any
testable, that is, refutable, content.
Rule no. 25, which demands that the modified theory 
``should pass some new, and severe, tests,'' is not (yet) satisfied.
And I would argue that rules no. 8 (``\textit{criteria of refutation} have to be laid down beforehand)
and no. 9 (``in the case of a threat to our system, we will not save it by any
kind of \textit{conventionalist stratagem}'') are also violated.
Indeed a pervasive feature of the standard-model literature is the conviction that any anomaly will, 
eventually, be explainable within the paradigm.\footnote{A striking example is the standard-model response to the remarkably correlated distribution of satellite galaxies around 
the Milky Way and the Andromeda galaxy, observations that have no, even remotely, plausible explanation under that model. Kroupa (\citeyear{Kroupa2016}, p. 557) documents the variety of 
 `conventionalist stratagems' adopted by standard-model cosmologists 
 in response to those observations and concludes,
``The [standard-model] community appears to have developed an unhealthy sense of simply ignoring or burying previously obtained results if these are highly inconsistent with the standard
model of cosmology.''
While MOND does not make a clear prediction here,
the observed correlations do not constitute a prima facie problem for Milgrom's theory (\citeauthor{Pawlowski2018}  \citeyear{Pawlowski2018}).} 
This attitude often conflicts with rule no. 13 as well, which forbids the appeal to 
``derivations to be discovered in the future.'' 

It is fair to say that rules no. 14 and15 \textit{are} satisfied since the number of additional axioms
(two in this case) is small.\footnote{On the other hand, one could reasonably take the point of view
that postulate SCM1 comprises a very \textit{large} number of independent postulates,
since the specification of the dark matter distribution around any single galaxy requires
 a 3d function, and furthermore a function that is different for every galaxy.}
But I would insist that rule no. 16 (``should proceed from some simple, new,
and powerful'' idea) is violated.
Here is one way to justify that statement: When teaching introductory astrophysics,
a problem that is commonly set to students (who have not yet learned about dark matter) 
is to ask them what can be inferred from the asymptotic flatness of rotation curves.
The `correct' answer, the answer that  students are expected to find, is:
there must be non-luminous matter in or around the galactic disk.
Far from being a ``new, and powerful'' idea, dark matter is almost literally the first
explanation that pops into anyone's head.

Finally, rule no. 26 recommends, when deciding between two theories, to choose
the theory that is ``testable in the most rigorous way.''
That recommendation would clearly favor Milgrom's theory over the standard model,
at least in terms of their postulates that target the rotation curve anomaly.

%%%%%%%%%%%%%%%%%%%%%%%%%%%%%%%%%%%%
\bigskip
\centerline{\textsection\textsection}
\bigskip	
%%%%%%%%%%%%%%%%%%%%%%%%%%%%%%%%%%%%

Rule no. 7 (Popper \citeyear{LSD}, p. 56) states
\begin{description}
\item only such statements may be introduced in science as are inter-subjectively testable 
\end{description}
and rule no. 24 (Popper \citeyear{LSD}, p. 268) is
\begin{description}
\item we shall not continue to accord a positive degree of corroboration to a theory 
which has been falsified by an inter-subjectively testable experiment.
\end{description}
Of course, as Popper acknowledged, and as others (Kuhn, Lakatos, Feryerabend) 
also emphasized, a falsifying instance -- even an inter-subjectively accepted one -- 
need not signal the ultimate death of a theory;
indeed many of the rules in Table~\ref{Table:Rules} are guidelines for the scientist
seeking to \textit{modify} her theory in response to a falsification.

But a theory that is not testable is not falsifiable.
Nevertheless, standard-model cosmologists do acknowledge that their theory
is inconsistent with a number of well-established facts.
\cite{SilkMamon2012} list seventeen such inconsistencies; 
\cite{Bullock2017} list about a dozen; and \cite{Kroupa2012}
gives twenty-two.
Which, if any, of these instances constitute falsifications in the sense that Popper 
used that term?

I would argue that there have been only two important instances (since the 1960s) 
where the
standard cosmological model has made predictions that were inter-subjectively testable; 
and that in both cases, the predictions were subsequently contradicted by observations.

The first instance pre-dated the dark matter postulates: it was the demonstration that
galaxy rotation curves are not correctly predicted by Newton's theory
 (Rubin et al. \citeyear{Rubin1978}; Bosma \citeyear{Bosma1981}).

The second occurred twenty years later: the discovery that the 
cosmological expansion is accelerating rather than decelerating, as Einstein's
equations generically predict  (Riess et al. \citeyear{Riess1998}; Perlmutter et al. \citeyear{Perlmutter1999}).

One feature that sets these two failures of prediction apart is the
way  the standard-model community chose to respond to them.
In both cases,  a (nearly) immediate and (nearly) unanimous consensus was
reached that an auxiliary hypothesis should be added to the theory: 
`dark matter' (that is, SCM1) in the first case, `dark energy' in the second
(Longair \citeyear{Longair2006}).
These two hypotheses were designed to maintain the integrity of Einstein's (or Newton's) 
theory of gravity in the face of  falsifying data, and indeed the
assumed properties of both dark matter and dark energy have been revised
a number of times, as needed to maintain that integrity as new data emerged
(Wang \citeyear{wang2010}; Majumdar \citeyear{Majumdar2015}).
Following \cite{Lakatos1978}, we can therefore identify Einstein's (Newton's) theory of gravity 
as constituting part of the `hard core' of the standard cosmological `research program,'
and the postulates relating to dark matter and dark energy as part of the
`protective belt' of auxiliary hypotheses that serve to maintain the integrity of that hard core
in the face of refutations.

By contrast, almost all of the other standard-model problems listed by
Silk \& Mamon, Bullock \& Boylan-Kolchin and Kroupa are failures of
\textit{accommodation}, not of \textit{prediction}.

What I mean by ``accommodation'' is best illustrated by an example.
Consider disk galaxy rotation curves.
As noted above, dark matter postulate SCM1 makes no prediction about rotation curves,
nor have standard-model cosmologists yet come up with any scheme that allows them
to predict the rotation curve of any galaxy.
However, standard-model theorists have devoted enormous effort to `getting rotation curves right':
that is: to finding ways to simulate the formation and evolution of galaxies starting from early times,
such that the \textit{statistical} properties of the \textit{simulated} galaxies match 
(in some specified sense)
the statistical properties of real galaxies -- ``properties'' here defined, of course,
in terms of the observable matter (stars, gas).
When a standard-model cosmologist says that he has failed to predict the rotation
curves of (say) dwarf galaxies, what he means is that he has not been able to find
a physically reasonable set of simulation parameters that results in simulated galaxies
whose structure and kinematics `look', in some average sense, like those of observed dwarf galaxies.

From a Milgromian perspective, these standard-model failures reflect the difficulty of
getting the dark matter in the simulations to behave `correctly' -- to distribute itself
around every galaxy so that its stars and gas respond to the total gravitational force
in a manner that mimics the modified dynamics.
Standard-model theorists, by contrast, consider the behavior of the dark matter
in their simulations to be unproblematic; the problem,
as seen by them, is to get those pesky \textit{baryons} to behave.
(E. g. \citeauthor{Bullock2017} (\citeyear{Bullock2017}, p. 380): ``Within the standard $\Lambda$CDM model, most properties of small-scale structure can be modeled with high precision in the limit that baryonic physics is unimportant.'' ``Small-scale'' here refers to single galaxies.)

Nowhere are the standard-model failures of accommodation more striking than in
the case of dwarf galaxies, which (they would say) are `dark-matter dominated':
that is: fully in the Milgromian regime.
And it is the dwarf galaxy literature that provides some of the starkest illustrations
of just how far standard-model cosmologists have strayed from methodological 
rule no. 7, that ``only such statements may be introduced in science as are inter-subjectively 
testable'':

Standard-model cosmologists identify the retinue of observed dwarf galaxies orbiting
the Milky Way with the dark matter `sub-halos' that form in their simulations.
A problem immediately arises: the number of sub-halos in the simulations far
exceeds the number of observed satellite galaxies 
(e.g. Silk and Mamon (\citeyear{SilkMamon2012}, 939): ``The excessive
predicted numbers of dwarf galaxies are [sic] one of the most cited problems 
with $\Lambda$CDM. The discrepancy amounts to two orders of magnitude.'')
This discrepancy is called by standard-model cosmologists the `missing-satellites
problem,'\footnote{Terminology like this should be of interest to 
social epistemologists: it suggests that standard-model cosmologists, when
conceptualizing the physical world, privilege their simulations over the actual data.
The name that \textit{Milgromian} researchers attach to this standard-model
failure is the `dwarf over-prediction problem.' 
Milgromian researchers postulate a different origin for the satellite galaxies -- see
Kroupa \citeyearpar{Kroupa2012} -- and the small number of satellites observed
around the Milky Way constitutes no problem for them.}
and most attempts to solve it invoke some mechanism for heating or removing
gas from the sub-halos before the epoch of star formation. 
No single mechanism `works' across the full spectrum of dwarf galaxy
(that is, sub-halo) masses, and standard-model cosmologists blithely invoke
different mechanisms, as the need arises, to explain the data on different mass scales.
For instance, Bullock (\citeyear{Bullock2010}, p. 12), 
after listing the various mechanisms that
have been proposed for suppressing star formation in the sub-halos, remarks:
\begin{quotation}
each imposes a different mass scale of relevance \ldots If, for example, we found 
evidence for very low-mass dwarf galaxies $V_\mathrm{max}\sim5$ km s$^{-1}$ 
then these [galaxies] would be excellent candidates for primordial $H_2$ cooling `fossils' of 
reionization in the halo.
\end{quotation}
And Bullock and Boylan-Kolchin (2017, 370) write that
``while many independent groups are now obtaining similar results in cosmological
simulations of dwarf galaxies . . . this is not an ab initio $\Lambda$CDM prediction, and it
depends on various adopted parameters in galaxy formation modeling.''

As Karl Popper (\citeyear{RAS}, p. 168) remarked in his critique of Freud's theory,
``\textit{every conceivable case will become a verifying instance}'' (italics his).

%%%%%%%%%%%%%%%%%%%%%%%%%%%%%%%%%%%%
\bigskip
\centerline{\textsection\textsection}
\bigskip	
%%%%%%%%%%%%%%%%%%%%%%%%%%%%%%%%%%%%

Rules no. 11, 18, 19, 25, 26 and 29 direct the scientist to prefer theories with
greater explanatory power, content, or testability.
Popper considered these qualities to be closely linked,
and in \textit{Conjectures and Refutations} (p. 217) he called the 
requirement that theories evolve in the direction of increasing content
 the criterion of ``potential satisfactoriness'' or ``potential progress''.
He  defined a criterion of \textit{actual} scientific progress 
in terms of rules no. 25 and 26: that is: the requirement that
at least some of a theory's new content be experimentally confirmed
(p. 220).
As is well known, Lakatos \citeyearpar{Lakatos1978} followed Popper's lead
in defining the ``empirical progressivity'' of an evolving theory
in terms of corroborated excess content -- that is: confirmed, novel predictions.

Popper (\citeyear{Popper1962}) sought to strengthen his intuitive idea of progress by
defining the  ``verisimilitude'' or ``truthlikeness'' of a theory, a measure of the theory's closeness to truth.
Given some definition of verisimilitude, and some scheme for evaluating it,
progress could be identified with an increase in a theory's verisimilitude.
But Popper did not view verisimilitude as a concept that should necessarily 
take the place of, or transcend, methodological considerations:
\begin{quotation}
I do not suggest that the explicit introduction of the idea of verisimilitude will lead to 
any changes in the theory of method. On the contrary, I think that my theory of testability
or corroboration by empirical tests is the proper methodological counterpart to this new
metalogical idea. The only improvement is one of clarification
(Popper \citeyear{CR}, p. 235).
\end{quotation}
(In Watkins's (\citeyear{Watkins1978}, p. 365) words: 
``Popper got along well enough without the idea of verisimilitude
for a quarter century after 1934.'')
Nevertheless  Popper's ``new metalogical idea'' has been enthusiastically taken up 
by a generation of realist philosophers, including \citeauthor{Oddie1986} (\citeyear{Oddie1986}), \citeauthor{Niiniluoto1987} (\citeyear{Niiniluoto1987}), \citeauthor{Kieseppa1996} (\citeyear{Kieseppa1996}), 
\citeauthor{Kuipers2000} (\citeyear{Kuipers2000}), 
 \citeauthor{Zwart2011} (\citeyear{Zwart2011}) and others.

In view of this impressive body of work, it is natural to ask whether it is now feasible
to compare the two theories of cosmology in terms of their verisimilitude. 
I will argue that the answer is `no' and furthermore that the  
prospects for doing so in the future are bleak.

My first point  has to do with the way that scientific theories (including Milgrom's)
typically evolve.
As Feyerabend (\citeyear{AgainstMethod4}, p. 157) noticed,
\begin{quotation}
Theories which effect the overthrow of a comprehensive and well-entrenched
point of view. . . are initially restricted to a fairly narrow domain of facts, to a
series of paradigmatic phenomena which lend them support, and they are only
slowly extended to other areas.
\end{quotation}
Milgrom's theory in its current state is marvelously
successful at making novel predictions for galaxies and groups of galaxies, 
and it has also had some notable successes in anticipating  large-scale data 
(Merritt \citeyear{Merritt2020}, chapter 6).
But there is general agreement even among Milgromian theorists that a suitable, 
general relativistic version of the theory (or its equivalent) is not yet available;\footnote{That statement
was written in 2019 and is no longer correct; see C. Skordis and T.  Złośnik, New relativistic theory 
for modified Newtonian dynamics, \textit{Physical Review Letters} 127(16), article id.161302.}
in the language of Imre Lakatos, the Milgromian research program is in
an earlier stage of development than the standard cosmological research program.\footnote{This
difference reflects the enormous disparity in number of scientists working in the two
research programs, as well as the disinclination of government agencies to fund Milgromian 
researchers, among other possible factors.}
And so comparisons with standard-model explanations of 
large-scale structure or the high-redshift universe would be pointless and,
in all likelihood, misleading.
 
 One might hope to sidestep this difficulty by comparing the verisimilitude of the
 two theories only in some narrow regime where both claim to make predictions; for instance,
 the internal kinematics of dwarf galaxies (e.g. Lazutkina (\citeyear{LazutkinaThesis}) who applies
 Niiniluoto's (\citeyear{Niiniluoto1999}) measure of truthlikeness to velocity data for
 a set of dwarfs).
 But such a project runs solidly up against an intractable problem.\footnote{The 
difficulty discussed in this paragraph exists  
 for any criterion of success that is essentially empirical or instrumentalist, e.g.
 Carnap's (\citeyear{Carnap1950}) `qualified instance confirmation,'
 Laudan's (\citeyear{Laudan1978}) `problem-solving efficiency,'
 van Fraassen's (\citeyear{Fraassen1980}) `constructive empiricism' etc.}
 Milgrom's theory is quite capable of making inter-subjectively testable predictions about galaxies.
 The standard cosmological model -- due largely to the vagueness of the dark matter hypothesis --
 is not; as we have seen, the best it can hope for is to include a scheme for simulating galaxy evolution
 that leads to model galaxies that look, in some average or statistical sense, 
 like real galaxies.
 But even standard-model cosmologists can be quite frank about the degree to which
 their numerical experiments are \textit{explicitly designed  to reproduce known facts}.
 For instance, one researcher writes\footnote{Quoted by Merritt (\citeyear{Merritt2020}, p. 75) who
 gives the source. ``Sub-grid models'' refers to algorithms that are meant to represent, 
 in some approximate manner, physical processes that occur on scales of time or
 space that are far too small to be simulated directly, e.g. turbulence, stellar winds etc.}
 \begin{quotation}
 Galaxy formation simulations \dots tune parameters such that the simulations produce
realistic-looking galaxy populations. In this sense the sub-grid models are `validated' as 
 `realistic' models by plausibility arguments in comparison to observations. 
Historically these models result from trial and error experiments. 
The models themselves might easily be `wrong' (in a strict physical sense) 
or assuming unrealistically high values for the coupling efficiencies -- they 
still produce realistic galaxy properties and the authors claim success.
\end{quotation}
The `success' of standard-model cosmologists at explaining
observations of galaxies is a function of a host of factors that are
external to their theory: 
how creative they were in crafting the ``sub-grid models'';
how effectively they were able to convince the larger community (including,
most importantly, journal referees and editors) of the 
physicality of those models;
how much time (human and computer) was available for the
simulations and their analysis (and, therefore, how much funding was available); etc.
As Thomas Kuhn might have said, these are tests of the theorist, not of the theory.

Given the fundamentally different ways in which the two groups of cosmologists achieve
correspondence of their theory with the facts, it is reasonable to ask whether 
case studies of methodology 
might not be a better guide to the progress-toward-truth of their respective
theories than measures of verisimilitude.
After all, one need not have a perfect criterion of justice (say) to know that there are certain
methodologies (e.g. deposition of witnesses) that are more conducive 
 to \textit{achieving} justice than others (e.g. divination).
In the same way, it is hard to believe that a critical, or falsificationist, approach to theory testing 
is less likely to lead to true theories than an uncritical, or verificationist, approach
(e. g. Agassi \citeyear{Agassi1959}; Popper \citeyear{Popper1962};
Albert \citeyear{Albert1987}; Gadenne \citeyear{Gadenne2006B}).

%%%%%%%%%%%%%%%%%%%%%%%%%%%%%%%%%%%%
\bigskip
\centerline{\textsection\textsection}
\bigskip	
%%%%%%%%%%%%%%%%%%%%%%%%%%%%%%%%%%%%

Niiniluoto (\citeyear{Niiniluoto1999}, p. 17) speculates about why 
a scientist would choose to follow a methodology like Popper's:
\begin{quotation}
\ldots for centuries, theory and practice
have already been in a mutual interaction in the field of scientific
inference. Scientists learn to do science through implicit indoctrination and
explicit instruction from their masters, textbooks, and colleagues. So if a
case study reveals that a group of real scientists favours `bold hypotheses'
and `severe tests', we may judge that they, or their teachers, have read
Popper. 
\end{quotation}
I am quite certain that Niiniluoto is mistaken here.
First of all, he is crediting  philosophers with far too much influence.
Most scientists -- particularly 
young scientists, but also the scientists who write the textbooks -- are dismissive
of philosophy, not to say contemptuous of it.
Scientists have reasons for doing the things they do, of course, but they don't get 
those reasons from the philosophers.

More to the point: Niiniluoto's explanation would imply that only \textit{Milgromian}
researchers have been brought up as critical rationalists, while the bulk of cosmologists
have been ``indoctrinated'' into some other epistemological school 
(Niiniluoto's inductivism, perhaps).
And \textit{that} hypothesis is easily debunked: The number of Milgromian researchers
is quite small (perhaps two dozen worldwide, certainly not many more); 
I know most of them personally; and
I can attest that their educations were quite of a piece with the educations of
the standard-model cosmologists in their cohorts.
There exists no secret society that is indoctrinating selected young scientists
into the Popperian mysteries.\footnote{Niiniluoto's ``implicit indoctrination'' calls to mind
 Tolstoy's (\citeyear{Tolstoy1903}) invocation of ``epidemic suggestions'' to explain the 
(to him) unfathomable popularity of Shakespeare's plays.}

Here is what does impress a scientist: a bold new conjecture that bears fruit.
The paradigmatic example, one that every physical scientist learns about early in their education,
is the set of postulates from which Bohr derived the energy levels of the hydrogen atom.
Bohr's success is impressive because it was so improbable.
Einstein (speaking at a time when Popper was eleven years old) declared that ``There must be something behind it. I do not believe that the derivation of the absolute value of the Rydberg constant is purely fortuitous.''\footnote{Quoted by Jammer
(\citeyear{Jammer}, p. 86). Jammer gives the original German in his note 107 as
``da mu\ss\ etwas dahinter sein; ich glaube nicht, da\ss\ die Rydbergkonstante durch Zufall in absoluten Werten ausgedr\"uckt richtig herauskommt.'' }
And it is obvious to any beginning student of quantum mechanics that a `turn-the-crank'
methodology like abduction or inference-to-the-best-explanation
could not possibly have led Bohr to his bold conjecture, a conjecture that went far
beyond the evidence that motivated it.

I am sure that standard-model cosmologists are just as impressed as other scientists
by instances in which a bold hypothesis survives a severe test.
But the standard cosmological model (at least since the addition of dark matter ca. 1980)
is simply not suited to making testable predictions, much less bold ones.
So standard-model cosmologists have, understandably, resigned themselves to the post hoc accommodation of new data, typically via large-scale computer simulations, and typically only in a
statistical sense.
Whereas Milgrom's bold theory \textit{is} eminently testable, even using data from a single
galaxy, and (as we have seen) its
novel predictions have again and again survived attempts to refute them.
One need look no farther to understand why Milgromian researchers
have stuck with a methodology that aligns with Popper's.

%%%%%%%%%%%%%%%%%%%%%%%%%%%%%%%%%%%%%%%%%%%%%%
%%%%%%%%%%%%%%%%%%%%%%%%%%%%%%%%%%%%%%%%%%%%%%
%\begin{thebibliography}{99.}
%\end{thebibliography}

\bibliography{bibliography}\label{refs}

\end{document}